\documentclass[sigconf]{acmart}
\begin{document}

\author{Jiarui Chen}
\affiliation{%
 \institution{National University of Singapore}
 \country{Singapore}
}
\email{e0893429@u.nus.edu}
\begin{abstract}
  Large language models (LLMs) have demonstrated significant potential in solving recommendation tasks. With proven capabilities in understanding user preferences, LLM personalization has emerged as a critical area for providing tailored responses to individuals. Current studies explore personalization through prompt design and fine-tuning, paving the way for further research in personalized LLMs. However, existing approaches are either costly and inefficient in capturing diverse user preferences or fail to account for timely updates to user history. To address these gaps, we propose the Memory-Assisted Personalized LLM (MAP). Through user interactions, we first create a history profile for each user, capturing their preferences, such as ratings for historical items. During recommendation, we extract relevant memory based on similarity, which is then incorporated into the prompts to enhance personalized recommendations. In our experiments, we define a new task that enables testing with varying memory size under two scenarios: single domain where memory and tasks are from the same category and cross-domain (e.g. memory from movies and recommendation tasks in books). The results show that MAP outperforms regular LLM-based recommenders that integrate user history directly through prompt design. Moreover, as user history grows, MAP's advantage increases in both scenarios, making it more suitable for addressing successive personalized user requests. 
\end{abstract}

\title{Memory Assisted LLM for Personalized Recommendation System}
\maketitle


\section{Introduction}

LLMs exhibit strong zero-shot learning abilities in traditional machine learning tasks, including recommendation systems (RS). In RS, LLMs generate recommendations from user-item interaction data framed as prompts \cite{gao2023chat, di2023retrieval}. However, these "one-size-fits-all" LLMs have proven insufficient for addressing personalized individual requests \cite{chen2024large}. As a result, there is growing interest in personalizing LLMs using personal data like behavioral histories to generate tailored responses \cite{tan2023user, eapen2023personalization}. By learning user preferences, LLMs can enhance performance in RS tasks to meet the increasingly customized needs \cite{tsai2024leveraging, kang2023llms}.

Researchers have explored personalized LLMs through two main approaches: fine-tuning and prompt design. Fine-tuning methods adjust LLM parameters based on user histories \cite{tan2024democratizing}, but face significant challenges including massive storage requirements for millions of user histories as training data and computational costs during fine-tuning \cite{salemi2023lamp}, along with issues like catastrophic forgetting \cite{wozniak2024personalized}.
Prompt-based approaches incorporating user history have proven more efficient, particularly when enhanced with strategies like history filtering and few-shot prompting \cite{dai2023uncovering, wang2023zero, salemi2023lamp, zhiyuli2023bookgpt}. However, these methods fail to account for dynamic user history sizes and exhibit inefficiencies when learning from extended user histories.

To address the limitations of existing research, we propose the \textbf{M}emory \textbf{A}ssisted LLM-based \textbf{P}ersonalized recommendation system (MAP), an approach that leverages prompt design in zero-shot settings, allowing users to continuously update their behavioral history with each new interaction. This design enables LLM-based recommendation systems to capture temporal changes in user behavior, becoming progressively more user-centric as the interactions between the individual and the system increase. Additionally, MAP incorporates a history retrieval mechanism that filters the user history to extract a "memory" of the most relevant past interactions.


For evaluation, we plan to test MAP in both single-domain and cross-domain recommendation scenarios. In the single-domain case, we will focus on movies, while in the cross-domain setting, we will use books as the recommendation target and movies as the cross-domain history. However, since existing benchmarks (e.g., LaMP) does not account for the effect of varying user history sizes on recommendation systems, we have defined a new task based on one of LaMP benchmark's existing tasks, LaMP-3 Personalized Product Rating \cite{salemi2023lamp}, to assess MAP with an increasing user history. Following prior research \cite{gao2023chat, kang2023llms}, we treat item rating prediction as a multi-class classification task.
In this new task, we conduct rating predictions iteratively, varying only two aspects: the size of the user history and the target of the prediction task. This approach allows us to investigate whether increasing user history enhances the system’s prediction accuracy. For cross-domain item rating prediction, we follow a similar iterative process, with the prediction target remaining the same for each user in every round. The objective is to determine if MAP's prediction accuracy for book ratings improves as more movie rating history is incorporated.
Experimental results show that the MAP framework enhances recommendation system performance, with improvements becoming more significant over time. This trend holds not only in single-domain item recommendations but also in cross-domain scenarios, further demonstrating MAP's effectiveness in long-term personalized recommendation tasks.

The main contributions of our work are as follows:
\begin{itemize}
    \item We propose a new method named MAP that dynamically integrates user interaction history to complete recommendation tasks with efficiency.
    \item We develop a new task from a popular benchmark LaMP \cite{salemi2023lamp} to enable evaluation with varying user history size and historical contexts.
    \item Our experiments show that MAP enhances RS performance in both single- and cross-domain scenarios, with greater improvements under growing memory size (e.g., in single domain, with 4 user history, MAP outruns the baseline for 0.039 mean absolute error (MAE); with 16 user history, MAP outruns the baseline for 0.104 MAE).
\end{itemize}


\vspace{-0.5em}
\section{Related work}
\subsection{Fine-tuning and Prompting}
After early studies have shown that LLMs can outperform popular recommendation algorithms, current work on LLM-powered RS is mainly from fine-tuning and prompting paradigms \cite{gao2023chat, kim2024large, zhao2024recommender}. However, fine-tuning is also widely recognized as computationally expensive and time-consuming, even when employing more efficient methods like Parameter-Efficient Fine-Tuning (PEFT) \cite{wozniak2024personalized}. On the other hand, the direct prompting paradigm for recommendation systems is more flexible and lightweight, as it bypasses the need for extensive model updates, offering a more efficient alternative for generating recommendations \cite{zhao2024recommender}.

\subsection{Personalization in Large language Model}
Current LLM personalization research primarily focuses on incorporating user-interaction histories through prompt design \cite{lyu2023llm, dai2023uncovering} and domain-specific knowledge integration \cite{yao2023knowledge, lian2024recai}. To improve efficiency, studies have explored long-term memory mechanisms, such as entity-based memory banks that store user attitudes from dialogue sessions \cite{xi2024memocrs}. However, these approaches lack relevant retrieval of user data, leading to low efficiency with long history contexts.

Retrieval Augmentation methods address this by filtering user behavior data before prompting \cite{salemi2024optimization}, evaluated using the LaMP benchmark with seven personalized tasks \cite{salemi2023lamp}. However, LaMP lacks evaluation across varying user history sizes, limiting assessment of personalized LLMs in different contexts. Recent frameworks have introduced optimization techniques like sequential sentiment labeling, user embedding generation, and recommendation simulation \cite{lee2025sealr, ning2025user, xi2024towards, zhang2024generative}, but these require costly fine-tuning compared to direct prompting approaches, use offline datasets for implementation, or result in an unscalable memory units. 

In this work, the proposed MAP furthers the study of personalized LLM-powered RS in user and item indexing and data augmentation according to suggestions of previous works on future directions \cite{zhao2024recommender}.

\vspace{-0.5em}
\section{Problem Statement}
\vspace{-0.5em}
In this section, we present the problem statement and describe the two personalization settings our study focuses on.

In the single-domain setting, each user provides a history composed of a sequence of item ratings, $\left(H_1, H_2, H_3, ..., H_n\right)$, where each rating is represented as a $\{(item, rating)\}$ pair. During the experiment, at iteration \textit{r}, the model predicts the rating for the $(\textit{r} + 1)$th item based on the previous \textit{r} item ratings. For instance, at the beginning of the experiment, the model is tasked with predicting the rating of $H_2$ given the item in $H_2$ and the history $\left(H_1\right)$. In the next iteration, this history expands to $\left(H_1, H_2\right)$, and the model then predicts the rating for $H_3$. Consequently, for each user, the model generates a sequence of predictions $\left(P_1, P_2, P_3, ..., P_{(n-1)}\right)$, where each prediction $P_i$ is dependent on the history $\left(H_1, ..., H_{(i-1)}\right)$.


In the cross-domain setting, each user has a history of item ratings $\left(H_1, H_2, H_3, ..., H_n\right)$ in domain $A$, and this history is used to predict ratings for items in domain $B$. At each iteration, the model predicts the rating for a specific item in domain $B$, producing a prediction $P_i$ based on a subset of the history $\left(H_1, ..., H_i\right)$ for domain $A$. Thus, the final output for each user in the cross-domain setting is a sequence of predictions $\left(P_1, P_2, P_3, ..., P_n\right)$.


\begin{figure*}[t]
  \centering
  \includegraphics[width=0.8\linewidth]{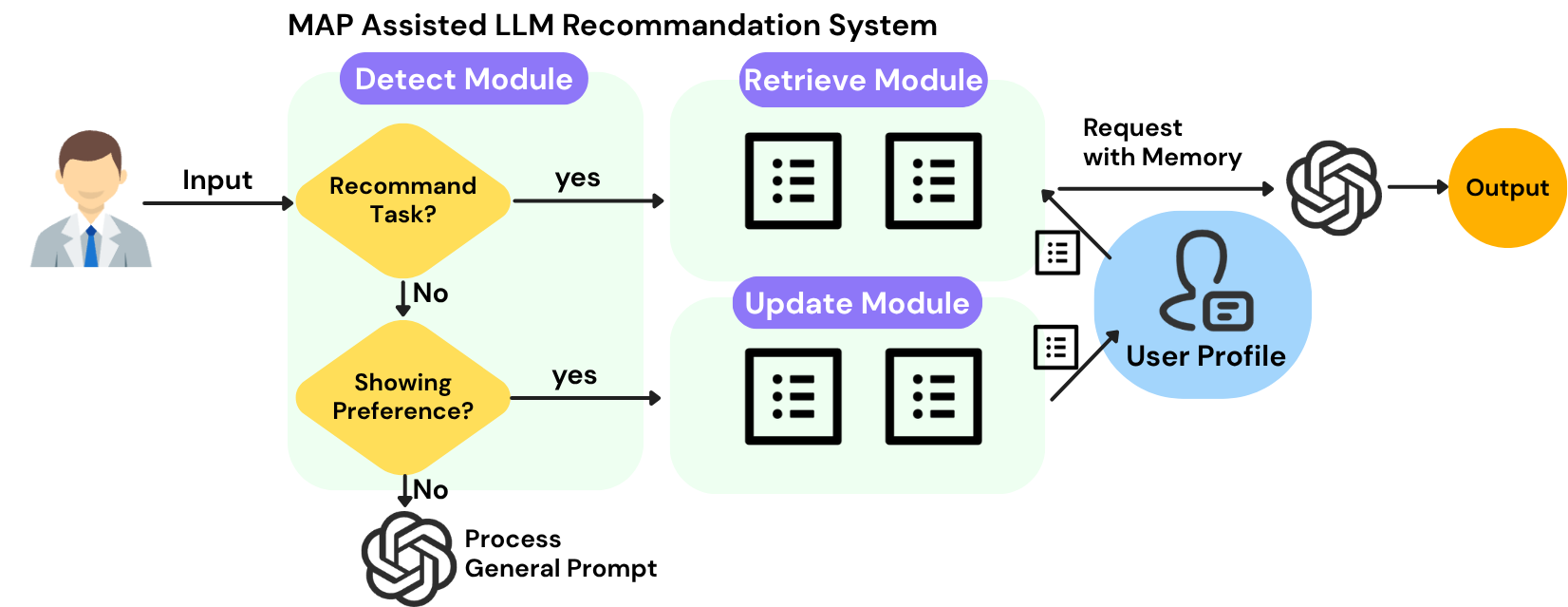}
  \caption{Pipeline of our proposed Memory-Assisted LLM-based recommendation system for personalization. The system consists of five key components: the detection module, user profile, retrieval module, update module, and the language model. The detection module classifies user queries, triggering either memory retrieval for generating recommendations or profile updates based on user preferences. The retrieved memory data is then passed to the language model to provide personalized recommendations.}
  \label{fig:pipeline}
\end{figure*}

\vspace{-0.5em}
\section{Method: Memory Assisted LLM-based Recommendation System}

In this section, we present the pipeline of the proposed Memory-Assisted LLM for Personalized Recommendation System (MAP). The flow of this system is illustrated in Fig.~\ref{fig:pipeline}. The core idea of memory assistance is to maintain a retrievable user profile, stored separately from the language model, functioning as a dynamic memory. This memory is continuously updated based on user interactions and retrieved to refine recommendation outputs by incorporating relevant user preferences. The primary objective is to enable recommendation systems (RSs) to generate more personalized responses tailored to each individual user.
The MAP system consists of five key components: the user profile, detection module, retrieval module, update module, and a basic large language model. The user profile stores information on the user's historical interactions and preferences. The detection module monitors user interactions to identify changes or new inputs. The retrieval module accesses relevant portions of the user profile to be used for predictions. The update module ensures the memory is kept current by integrating new data after each interaction. Finally, the base large language model processes the inputs and generates personalized recommendations, enhanced by the retrieved user history.


\subsection{User Profiles}

The user profile serves as the system's memory storage, maintaining a comprehensive and categorized collection of data that reflects each user's preferences across various domains. This data includes movie ratings, product reviews, and other interactions, providing a detailed historical record that allows the system to access relevant user preferences whenever a recommendation or prediction is needed.


For each user, this structured memory is organized in a table format. In the case of movie ratings, for example, the table may include columns such as movie title, genre, date watched, and the user's rating. Each row represents a distinct interaction, recording the user's experience with a specific movie. Similar tables are maintained for other domains, such as product reviews or book ratings, ensuring that user preferences are stored systematically.


This structured format allows the system to efficiently retrieve specific entries relevant to a given query, enabling it to concentrate on the user's preferences within a particular context or genre. This enhances the system's ability to provide highly personalized recommendations based on the user's past interactions.


\vspace{-0.5em}
\subsection{Detection Module}

Compared to the original LLM, our MAP system introduces two additional behaviors specifically tailored for recommendation tasks: (1) constructing or updating the memory, which serves as the user profile, and (2) retrieving this memory to generate personalized recommendations. All other general functionalities of the LLM remain unchanged. To handle these new behaviors, a detection module is employed to classify the type of action required for each incoming query during user interactions.


We categorize incoming queries into three types:

\begin{itemize}
\item  \textbf{Type A:} A request for a recommendation. This type requires retrieving the memory from user profile to generate personalized suggestions.
\item \textbf{Type B:} A message that reflects a new user preference, which prompts the system to update the user profile in memory with the new information.
\item \textbf{Type C:} A query unrelated to recommendations, where the LLM responds directly without utilizing the memory system.
\end{itemize}

\begin{figure}
  \centering
  \includegraphics[width=\linewidth]{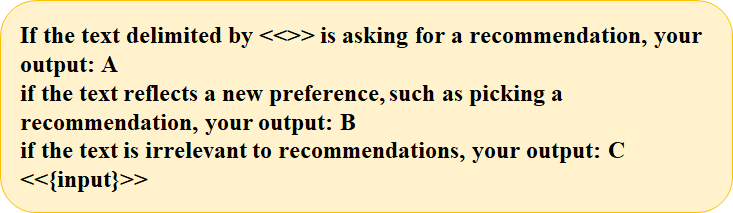}
  \caption{The prompt used for detect module.}
  \label{fig:detect}
\end{figure}

For instance, when a user asks for a movie recommendation, the system classifies this as a Type A query, since it must access the user's preferences, such as favorite movie genres or past ratings, to provide relevant suggestions. In this case, the user’s previous interactions, stored in the user profile, are passed to the retrieval module to assist in generating personalized recommendations.
On the other hand, if the user rates or selects a movie, this is classified as a Type B query, as it reflects a new preference that needs to be integrated into the user profile for future recommendations. This update ensures that the system continually refines its understanding of the user's tastes.
To manage this process, the detection module relies on a prompt template, as illustrated in Fig.~\ref{fig:detect}. For each incoming query, we prompt the LLM with additional steps to first identify the type of query—whether it is Type A, B, or C—before proceeding with the corresponding action. This ensures that the system correctly processes and responds to various user inputs, enhancing both the personalization and efficiency of the recommendation process.


\subsection{Memory retrieval and Memory Assisted Recommendation} 
\begin{figure}[t]
  \centering
  \includegraphics[width=\linewidth]{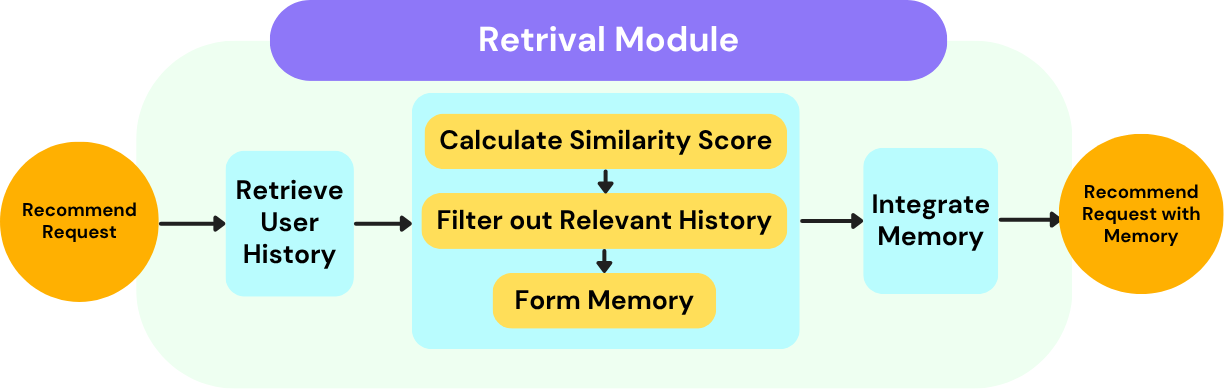}
  \caption{Inner Structure of the Retrieval Module. The retrieval module consists of three key steps: 1) loading the stored original user profiles; 2) calculating similarity scores between stored items and the current query, and 3) selecting the most relevant items from the user's history. These relevant items are then fed into the language model to assist in generating personalized recommendations.\vspace{-0.5em}}
  \label{fig:retrival}
\end{figure}

When a query involves generating a recommendation, the retrieve module processes the recommendation request by augmenting it with filted memory from the user profile. The work flow of the retrieve module is illustrated in Fig.~\ref{fig:retrival}. 

\paragraph{\textbf{Retrieve User History}}
The module first loads the user's stored history from memory, which typically consists of a list of items, including details such as the name, description, and historical rating. However, loading the entire user history every time a recommendation is requested can be inefficient, as many entries may be irrelevant to the current query. Moreover, large language models (LLMs) have inherent limitations, including high computational costs for processing long inputs and constraints on the maximum sequence length.
To address these challenges, we draw inspiration from previous studies that have shown the effectiveness of in-prompt retrieval augmentation for personalizing LLMs \cite{salemi2023lamp}. In our memory-assisted approach, rather than loading all user data, the retrieval module selectively extracts the most relevant data points from the user profile based on the current query. This strategy enables the system to focus on the most pertinent subset of data, improving both the efficiency of processing and the quality of the generated recommendations. By limiting the input to only the most relevant user preferences, the system enhances its ability to produce accurate, personalized recommendations without overwhelming the LLM with unnecessary data.

\paragraph{\textbf{Calculate Similarity Score \& Form Memory}}
Since each item in the user's history contains a detailed description, we use the similarity between the item that needs to be predicted and the historical items to filter and rank the retrieved memory. This similarity is computed using a scoring mechanism that compares the description and attributes of the item to be predicted with those of each item in the user's history. The similarity score for each historical item is calculated as follows:
\begin{equation}
    \text{Score}_i = \text{Similarity}(item_i, x)
\end{equation}
where $item_i$ represents each instance stored in memory, and $x$ is the input for the task requiring a rating prediction. The function $\text{Similarity}$ is a model that computes the similarity score between the historical items and the target item. By applying this scoring mechanism, the system can prioritize and retrieve the most relevant historical items, ensuring that only the most similar and relevant data points are used to inform the recommendation, thereby improving the quality and efficiency of the prediction process.


In our method, we explore different implementations for the similarity module depending on the type of description associated with each item. For items that include a genre list in their description, we calculate the similarity by comparing the intersection of genres between the predicted item and the historical items. This approach helps prioritize recommendations that align with the user’s preferences in terms of genre.


Beyond simple attribute matching, we also employ a more advanced technique using a pre-trained language model, such as BERT \cite{devlin2018bert}, for text feature extraction. By embedding the text descriptions of both the historical and predicted items, we compute the cosine similarity between their feature vectors. This allows for a semantic comparison, enabling the system to identify similarities between items even if they are described using different terms.

\begin{figure}
  \centering
  \includegraphics[width=\linewidth]{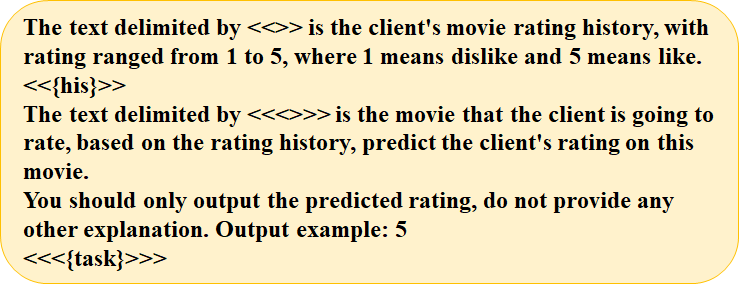}
  \caption{The prompt template which combines the retrieved memory and conducts recommendations for a new item.}
  \label{fig:prompt_retrival}
\end{figure}

\paragraph{\textbf{Integrate Memory}}
Once the most relevant items are identified and ranked, the processed memory is integrated into the LLM as part of the prompt. The template for the prompt is shown in Fig.~\ref{fig:prompt_retrival}. This augmented input enables the LLM to generate more personalized and contextually accurate recommendations by focusing on the user’s most relevant historical preferences. By narrowing the input to the most pertinent data, the model reduces computational costs while enhancing the quality of the predictions, ensuring a more efficient and personalized recommendation process.


\subsection{Memory update} 

The updating module integrates new data into the user profile whenever a new preference is detected from the input. This process is triggered when the system identifies a Type B query.

\begin{figure}
  \centering
  \includegraphics[width=\linewidth]{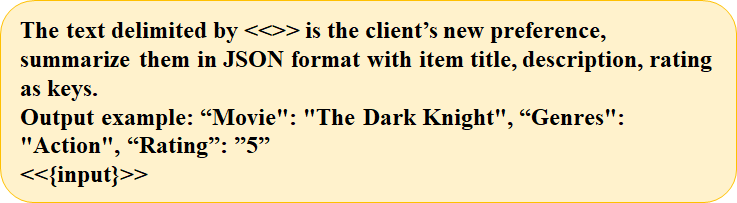}
  \caption{The prompt used for Update Module}
  \label{fig:update_prompt}
\end{figure}

When a user interacts with the system in ways that indicate new preferences—such as submitting a new rating or review—the system processes and stores this information within the user profile. The prompt template used for updating is shown in Fig.~\ref{fig:update_prompt}. The newly gathered data is structured in the same format as the existing entries in the user profile, maintaining consistency across all stored information.
This ensures that the system can seamlessly retrieve the updated data for future queries, keeping the user profile reflective of the most current and accurate preferences. For example, if a user rates a movie after watching it or leaves a review detailing their experience, the updating module records this information, including attributes such as the item's name, description, rating, and any additional context like genre or category tags. By capturing this data, the system becomes more capable of delivering personalized recommendations that reflect the user’s evolving preferences.


\vspace{-0.5em}
\section{Experiments}
In this section, we evaluate the effectiveness of our Memory-Assisted Personalized Recommendation System (MAP) through a series of experiments. The experiments are designed to assess the system's performance in two key scenarios:

\begin{itemize}
\item  \textbf{Single-domain movie rating prediction:} In this scenario, we evaluate how well MAP predicts movie ratings based on a user's past interactions within the movie domain. The system uses the user’s movie rating history to generate personalized recommendations for new movies.
\item \textbf{Cross-domain book rating prediction based on movie ratings:} In this scenario, we assess MAP's ability to predict book ratings by leveraging the user's movie rating history. This cross-domain evaluation examines whether MAP can effectively transfer learned user preferences from one domain (movies) to another (books), showcasing its versatility in handling multi-domain recommendation tasks.
\end{itemize}

By conducting these experiments, we aim to determine the impact of memory-assisted retrieval and personalized prompt design on recommendation accuracy and user satisfaction in both single-domain and cross-domain contexts.


\subsection{Experimental Details}
We use the GPT-3.5-turbo model \cite{gpt-turbo} for our study, as prior research has demonstrated the superiority of ChatGPT in similar recommendation tasks \cite{dai2023uncovering, di2023evaluating, silva2024leveraging}. Our study includes two experiments: a movie rating prediction experiment and a cross-domain rating prediction experiment. Both experiments utilize a retrieval approach to select relevant rating histories as memory to assist the LLM model in learning user preferences.


In the movie rating prediction experiment, we sort the user’s rating history based on the number of matching movie genres, extracting ratings for movies with the highest genre overlap as part of the prompt input for the model. This approach ensures that the memory provided to the LLM focuses on items most similar to the current prediction task, thereby improving the relevance of the personalized recommendations.


In the cross-domain prediction experiment, we select movies and books as the two domains for testing. The primary focus is to determine whether the accuracy of book rating predictions by the memory-assisted LLM improves with an increasing amount of movie rating history used as memory. Since genres differ between these domains, we utilize a pre-trained BERT tokenizer to encode the genres of both books and movies. We then calculate the cosine similarity between these encoded representations. Similar to the first experiment, the memory consists of ratings for movies with high cosine similarity to the book being predicted.


For evaluation, we use mean absolute error (MAE)—a commonly employed metric in rating prediction—to assess the performance in both experiments. This metric helps us quantify the accuracy of the system’s predictions by measuring the average magnitude of the errors between the predicted and actual ratings.

\vspace{-0.5em}
\subsection{Baseline Model}
Based on previous approach \cite{dai2023uncovering}, we developed a baseline model that directly prompts the LLM with the unprocessed rating history, without using extracted memory. This baseline model serves as a comparison for evaluating the effectiveness of our memory-assisted approach using the GPT-3.5-turbo model.


In the movie rating prediction task, the baseline model treats the user's past ratings as a series of historical messages. The model is prompted with a prediction task query and an updating prompt for each iteration. The updating prompt informs the model of the actual rating for that round's query. For subsequent iterations, the baseline model is given both the current query and the past queries along with their corresponding updating prompts, essentially providing the model with the user's historical interactions as messages. The baseline model also includes an additional "history prompt," ensuring that every starting iteration for each user begins with an identical input structure. This design simulates a scenario in which users repeatedly query a RS without the assistance of memory extraction or selection mechanisms. The model essentially recalls the entire interaction history as flat sequences of messages.


For cross-domain rating prediction, where the task is to predict book ratings based on movie ratings, we adjust the baseline model by removing the updating prompts from each iteration. This is because the prediction task (a single book rating) remains constant across iterations. Instead, the model receives one book as the prediction object and the user's movie rating history. In the baseline model, the user's movie rating history grows incrementally over iterations, with each rating presented as a separate sentence. Unlike the memory-assisted model, which filters and extracts the most relevant historical ratings for each iteration, the baseline model feeds one movie rating at a time into the LLM as historical messages alongside the task query for each iteration. This results in a more fragmented input structure compared to the memory-assisted model, where movie rating history and task queries are integrated into a complete prompt for each iteration.


\subsection{Single-domain Recommendations}
We first evaluate the single-domain recommendation task using movie rating prediction. In this scenario, each user is provided with a list of historical movie ratings reflecting their preferences, and the goal is to predict the next movie rating. For each user, the historical data starts with 1 entries, and with each iteration, additional rating history is incrementally added, growing from 1 to 18 entries.


\paragraph{\textbf{Dataset}}
The memory-assisted model is evaluated using the widely adopted MovieLens 100k dataset \cite{harper2015movielens}, which contains 944 users and 1,683 movies. To ensure sufficient data per user, we exclude users with fewer than 19 movie ratings and limit each user to exactly 19 ratings. After preprocessing, the dataset consists of 839 users and a total of 15,941 movie ratings. For evaluation, we compare the predicted rating for each user’s next movie with the corresponding ground truth rating.


\paragraph{\textbf{Results}}

\begin{figure}[!h]
  \centering
  \includegraphics[width=\linewidth]{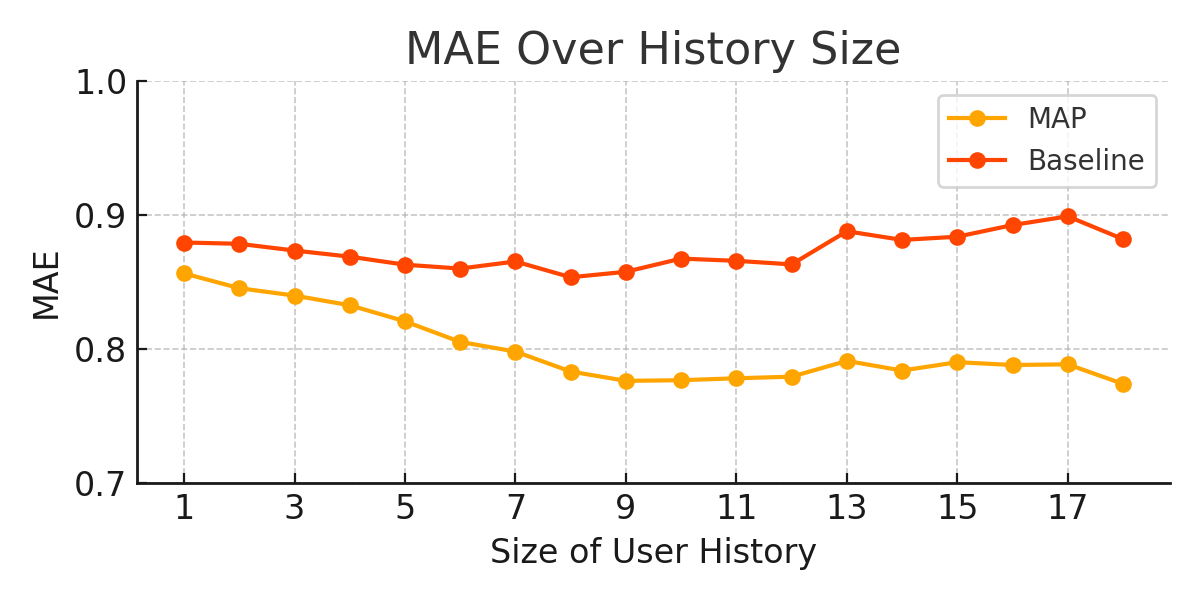}
  \caption{MAE trends over different user history sizes for the our MAP and baseline for the single-domain recommendation. Our MAP shows a consistent decline in MAE as the user history increases, while the baseline model maintains stable performance. }
  \label{fig:exp1}
\end{figure}

For each user, the model generates 18 predictions for 18 different movies. We compute the absolute difference between the predictions and the users' actual ratings and calculate the Mean Absolute Error (MAE) across the 834 users. From the visualization in Fig.~\ref{fig:exp1}, the memory-assisted model consistently outperforms the baseline starting from the second iteration, with similar performance to the baseline during the first iteration.


A notable trend is the decreasing MAE for the MAP as the amount of user history increases, while the baseline model's MAE remains relatively stable across all iterations. As shown in Table~\ref{tab:experiment1_results}, the memory-assisted model (MAP) provides an average improvement of 4.89\% in MAE when the user history size reaches 5 entries, and this improvement grows to 12.27\% when the user history size increases to 17 entries. This results in a widening gap between the two models' MAE values as the user's movie rating history grows. These findings indicate that incorporating memory enables the LLM-based recommendation system to deliver more accurate predictions as more user history is accumulated, significantly enhancing personalization.


\begin{table}[h]
\centering
\caption{Mean Absolute Error (MAE) of single-domain recommendation under different user history sizes}
\label{tab:experiment1_results}
\begin{tabular}{c|c|c|c|c}
\toprule
\textbf{User's history size} & 5 & 9 & 13 & 17 \\
\midrule
Vanilla GPT & 0.8628 & 0.8576 & 0.8878 & 0.8989 \\
\textbf{MAP (Ours)}  & \textbf{0.8206 }& \textbf{0.7763} & \textbf{0.7911} & \textbf{0.7886} \\
\bottomrule
\end{tabular}
\end{table}


\paragraph{\textbf{Costs}}

\begin{table}[h]
\centering
\caption{The average cost of running when increasing every 10 history for one user. This is estimated by the average token used for each piece of user history from MovieLens dataset and the cost of API for GPT model.}
\label{tab:experiment1_cost}
\begin{tabular}{c|c}
\toprule
\textbf{Method} & Cost of dollar / 10 history per user  \\
\midrule
Vanilla GPT & $\$0.00325$ \\
\textbf{MAP (Ours)}  &  $\textbf{\$0.00086}$ \\
\bottomrule
\end{tabular}
\end{table}

Beyond performance, we also compare the cost of running the recommendation models using the same API of the large language model (see Table~\ref{tab:experiment1_cost}). For both our model and the baseline model, we use the pricing of GPT-3.5-turbo \cite{openaipricing}. For the MovieLens dataset, we calculate the average number of tokens processed per user history and multiply this by the API's cost per token. The baseline model (Vanilla GPT) requires more tokens to generate recommendations due to its less efficient handling of user history, resulting in a higher operational cost of $\$0.00325$ per 10 histories per user. In contrast, our proposed MAP model optimizes token usage by summarizing and retrieving relevant user history, thereby reducing the cost to $\$0.00086$ per 10 histories per user. This significant cost reduction demonstrates the economic advantage of our approach, making it more scalable and sustainable for large-scale deployments.



\subsection{Cross-domain Recommendation}

To evaluate the model's ability to perform cross-domain recommendations, we designed an experiment where the task is to predict book ratings based on a user's movie rating history. This experiment tests the model’s ability to transfer knowledge from one domain (movies) to another (books) and still provide relevant recommendations. For each user, the experiment is conducted over 18 iterations, with the number of movie ratings used in each iteration incrementally increasing from 1 to 18. The memory-assisted model retrieves relevant movie ratings from the user's profile and calculates the cosine similarity between movie genres and book genres. The most similar movie ratings are used to assist in predicting the user’s book rating.


\paragraph{\textbf{Dataset}}

For this cross-domain experiment, we use movie and book ratings from the Amazon Review Data \cite{ni2019justifying}. We employ the "ratings only" version of the dataset and merge it with item metadata to ensure both movie and book ratings are aligned with their respective genres. After filtering out users with fewer than 18 movie interactions and capping the number of movie ratings at 18 per user, the refined dataset consists of 533 users, 533 book ratings, and 9,594 movie ratings.


\paragraph{\textbf{Results}} 

\begin{figure}
  \centering
  \includegraphics[width=\linewidth]{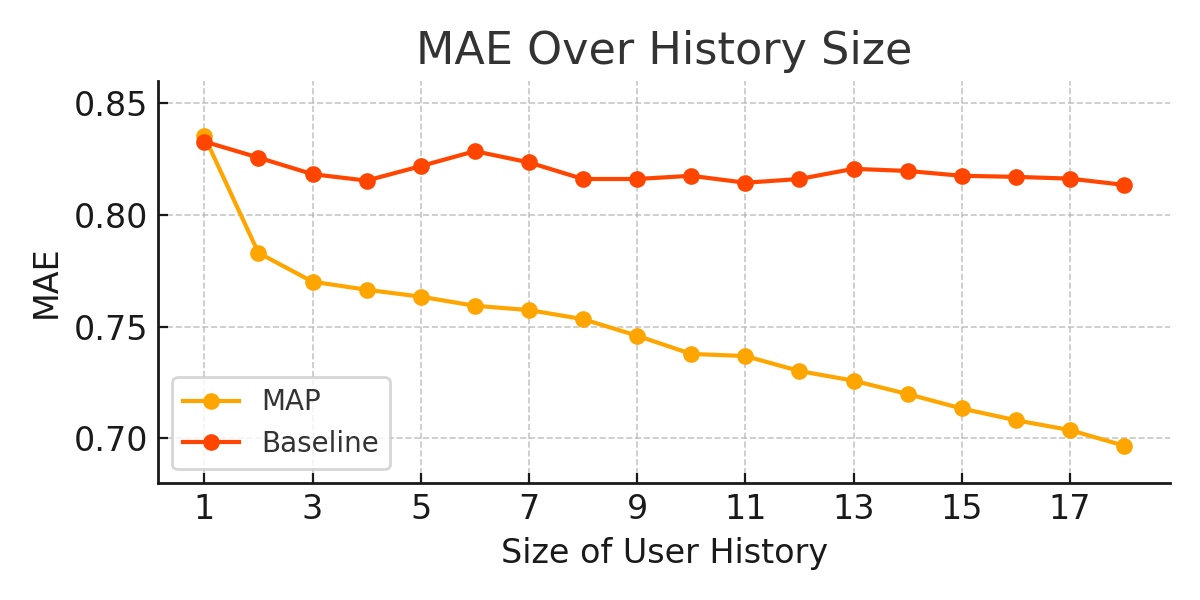}
  \caption{MAE results for cross-domain recommendation tasks. The memory-assisted model shows a clear improvement as the number of movie ratings increases, compared to the baseline model which struggles to maintain its initial performance as the user history grows.}
  \label{fig:exp2}
\end{figure}

\begin{table}[h]
\centering
\caption{Mean Absolute Error (MAE) of cross-domain recommendation under different user history sizes.}
\label{tab:experiment2_results}
\small
\begin{tabular}{c|c|c|c|c}
\toprule
\textbf{User's history size} & 5 &9 & 13 & 17 \\
\midrule
Vanilla GPT & 0.8218 & 0.8160 & 0.8206 & 0.8162 \\
\textbf{MAP (Ours) } &\textbf{ 0.7634} & \textbf{0.7459 }& \textbf{0.7258 }& \textbf{0.7037} \\
\bottomrule
\end{tabular}
\end{table}


For each user, the model generates 18 predictions for a single book, each based on an incrementally growing history of movie ratings ranging from 1 to 18 entries. The absolute difference between each predicted book rating and the actual rating is calculated, and the Mean Absolute Error (MAE) is computed across all users. To ensure robustness, we shuffle the order of the movie history for each user and repeat the experiment for both the memory-assisted model and the baseline model. The final MAE is averaged across the shuffled and unshuffled results to eliminate potential biases from the order of history inputs.


As shown in the smoothed graph in Fig.~\ref{fig:exp2}, the memory-assisted model consistently outperforms the baseline (Vanilla GPT) on average. The memory-assisted model shows significant improvement as more movie rating history becomes available, especially in later iterations. The MAE for the MAP steadily declines as the movie history grows, while the baseline model exhibits a temporary dip in error during the first five iterations but returns to its initial performance levels afterward. Table~\ref{tab:experiment2_results} provides more detailed values, showing that MAP achieves an average improvement of 7.10\% over the baseline when there are 5 user histories and an even higher average improvement of 13.78\% when the history size reaches 17 entries.


This performance suggests that the baseline model, which handles larger user history directly within its prompt, struggles with efficiently managing increasing amounts of input data, leading to a degradation in performance. In contrast, the memory-assisted model maintains its ability to personalize recommendations more effectively as the user history expands, even in a cross-domain setting, where movie ratings are used to predict book ratings.


\paragraph{\textbf{Costs}}

\begin{table}[h]
\centering
\caption{The average cost of running when increasing every 10 history for one user.}
\label{tab:experiment2_cost}
\small
\begin{tabular}{c|c}
\toprule
\textbf{Method} & Cost of dollar / 10 history per user  \\
\midrule
Vanilla GPT &  $\$0.00322$ \\
\textbf{MAP (Ours)}  & $\textbf{\$0.00156}$ \\
\bottomrule
\end{tabular}
\end{table}


In addition to performance improvements, the memory-assisted approach significantly reduces computational costs, as shown in Table~\ref{tab:experiment2_cost}. Costs were measured by calculating the average expense incurred when adding 10 additional pieces of user history data for each user.


The cost per user for the Vanilla GPT model increases substantially as more user history is added, averaging \$0.00322 per 10 history entries. In contrast, the memory-assisted model (MAP) reduces this cost to \$0.00156 per 10 history entries. This cost reduction is achieved by focusing only on the most relevant user history through memory retrieval, minimizing the data the LLM processes.


This cost-efficiency is especially valuable in large-scale applications that manage thousands or millions of users with substantial historical data. By using the memory-assisted approach, systems can maintain or improve performance while reducing the operational costs associated with recommendation tasks, making it a scalable and sustainable solution.


\vspace{-0.5em}
\section{Conclusion}
In the context of the rapid development of large language models (LLMs), increasing attention has been drawn to the field of personalized LLMs. In this paper, we introduced MAP (Memory-Assisted Personalized LLM), a model that integrates users' past behaviors while updating with new user requests, demonstrating its advantages in supporting long-term interactions between users and recommender systems in both single-category and cross-domain recommendation tasks. By adopting a retrieval framework to filter and prioritize the most relevant historical data, MAP effectively captures patterns in user preferences, understanding their likes and dislikes over time. To further evaluate MAP in a personalized setting with progressively increasing user history, we designed a new task based on the LaMP benchmark \cite{salemi2023lamp}, enabling evaluation based on varying user history sizes and contexts. The proposed MAP structure and the new evaluation task tap into new potentials for personalization of LLM and provide a framework that could guide future work. Our findings underscore the importance of memory-assisted architectures in improving both the accuracy and efficiency of personalized recommendations.


\bibliographystyle{ACM-Reference-Format}
\bibliography{jerry-bib}


\end{document}